\documentstyle[sprocl]{article}

\input{psfig.sty}
\def\be{\begin{equation}}
\def\ee{\end{equation}}
\def\bea{\begin{eqnarray}}
\def\eea{\end{eqnarray}}
\def\lesssim{\mathrel{\raise1.16pt\hbox{$<$}\kern-7.0pt 

\lower3.06pt\hbox{{$\scriptstyle \sim$}}}}         

\begin{document}


\title{SHAPELETS: A NEW METHOD TO MEASURE GALAXY SHAPES\footnote{To
appear in Procs. of the Workshop ``The Shapes of Galaxies and their
Halos'', Yale, May 2001.}}

\author{A. Refregier}

\address{Institute of Astronomy, Madingley Road, Cambridge CB3
OHA, UK; ar@ast.cam.ac.uk}

\author{T.-C. Chang}
\address{Dept. of Astronomy, 550 W. 120 street, Columbia University,
New York, \\ NY 10027, USA; tchang@astro.columbia.edu}

\author{D. J. Bacon\footnote{previous address: Institute of Astronomy,
Madingley Road, Cambridge CB3 OHA, UK.}}
\address{Institute for Astronomy, University of
Edinburgh, Blackford Hill,\\ Edinburgh EH9 3HJ, UK; djb@roe.ac.uk}


\maketitle\abstracts{We present a new approach to measure the
shapes of galaxies, a fundamental task in observational astronomy.
This approach is based on the decomposition of a galaxy image into
a series of orthogonal basis functions, or `shapelets'. Our choice
of basis functions, namely the Gauss-Hermite series, has a number
of remarkable properties under distortions, convolutions and
noise, which makes them particularly well suited for astrophysical
applications. In particular, we describe how they can be used to
measure the shear induced by weak gravitational lensing, with the
precision required for upcoming surveys. We also show how
shapelets can be used to reconstruct images from interferometric
observations. Other application of shapelets, such as image
compression, PSF deconvolution, de-projection and the study of
galaxy morphology, are also briefly discussed.}

\section{Introduction}

The measurement of galaxy shapes is a fundamental task in
observational astronomy. In these proceedings, we present a new
method for shape measurements described in detail in Refregier
(2001).  It is based on the linear decomposition of each galaxy
into a series of localised basis functions with different shapes,
which we call `Shapelets'. As a basis set, we choose the
Gauss-Hermite series, whose remarkable properties make it ideal
for astronomical applications. In particular, we summarise the
results of Refregier \& Bacon (2001) who showed how shapelets can
be used to measure weak lensing, with the precision required for
upcoming surveys. We also describe how shapelets can be used to
reconstruct images from interferometric data, as described by
Chang \& Refregier (2001).

We illustrate the method using images from the Hubble Space
Telescope and from the FIRST radio survey. Finally, we describe
several further applications of shapelets and discuss their
relevance in the context of astronomical applications requiring
high precision.

\section{Shapelet Method and Properties}
\begin{figure}[t]
\centerline{\psfig{figure=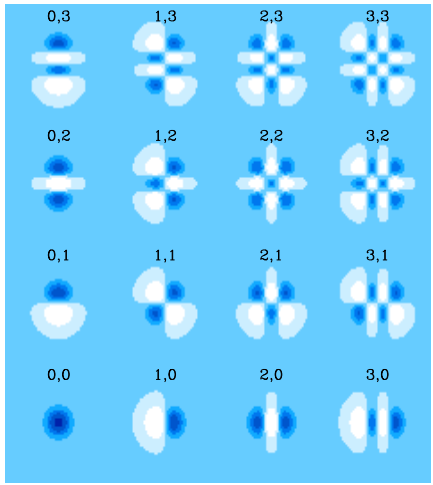,height=2.5in}} \caption{First
few 2-dimensional Cartesian basis functions $B_{n_{1},n_{2}}$. The
dark and light regions correspond to positive and negative values,
respectively. \label{fig:basis}}
\end{figure}

We begin by summarising the formalism of Refregier (2001) for a
description of galaxies in the Gauss-Hermite basis set. A galaxy
with intensity $f({\mathbf x})$ can be decomposed into our basis
functions $B_{\mathbf n}({\mathbf x};\beta)$ as
\begin{equation}
\label{eq:decompose} f({\mathbf x}) = \sum_{\mathbf n} f_{\mathbf
n} B_{\mathbf n}({\mathbf x};\beta),
\end{equation}
where ${\mathbf x}=(x_1,x_2)$ and ${\mathbf n}=(n_1,n_2)$.  The
2-dimensional cartesian basis functions can be written as
$B_{\mathbf n}({\mathbf x};\beta) = B_{n_1}(x_1;\beta)
B_{n_2}(x_2;\beta)$, in terms of the 1-dimensional basis functions
\begin{equation}
B_{n}(x;\beta) \equiv \left[ 2^{n}  \pi^{\frac{1}{2}} n! \beta
\right]^{-\frac{1}{2}} H_{n}\left(\frac{x}{\beta}\right)
e^{-\frac{x^2}{2 \beta^2}},
\end{equation}
where $H_{n}(x)$ is a Hermite polynomial of order $n$. The
parameter $\beta$ is a characteristic scale, which is typically
chosen to be close to the radius of the object. These basis
functions are the eigenstates of the Quantum Harmonic Oscillator
(QHO), allowing us to use the formalism developed for this
problem. Similar decomposition into basis functions has been
independently suggested by Bernstein \& Jarvis (2001). The first
few basis functions are shown in Figure~\ref{fig:basis}.

\begin{figure}[t]
\centerline{\psfig{figure=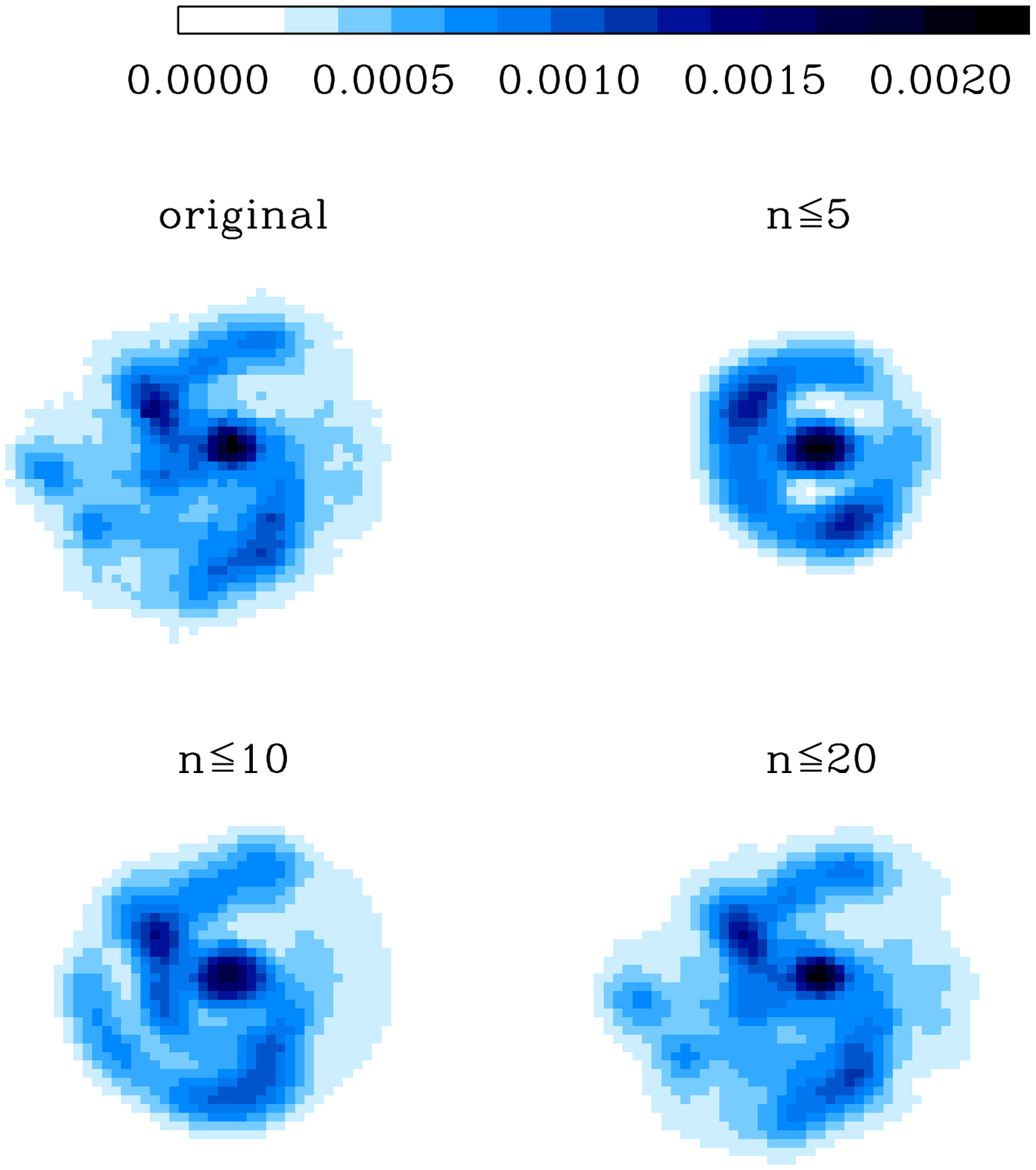,height=2.5in}}
\caption{Decomposition of a galaxy image found in the HDF. The
original $60\times60$ pixel HST image (upper left-hand panel) can
be compared with the reconstructed images with different maximum
order $n=n_{1}+n_{2}$. The shapelet scale is chosen to be
$\beta=4$ pixels. The lower right-hand panel ($n \leq 20$) is
virtually indistinguishable from the initial image.
\label{fig:galaxy}}
\end{figure}

Because these basis functions, or `shapelets', form a complete
orthonormal set, the coefficients $f_{\mathbf n}$ can be found
using
\begin{equation}
\label{eqn:decompose} f_{\mathbf n} = \int_{-\infty}^{\infty}
d^{2}x~f({\mathbf x})B_{\mathbf n}({\mathbf x};\beta).
\end{equation}
Figure~\ref{fig:galaxy} shows an example of the decomposition of a
galaxy in the Hubble Deep Field (Williams et al 1996). The details
of the image are fully reconstructed if coefficients up to $n=20$
are included; this decomposition thus provides an excellent and
efficient description of galaxy images in practice.

The chosen basis functions have a number of remarkable properties.
Firstly, they are invariant under Fourier transforms up to a
scaling factor, so that
\begin{equation} \label{eq:B_tilde}
\widetilde{B}_{n}(k;\beta) = i^{n} B_{n}(k;\beta^{-1}),
\end{equation}
where tildes denote Fourier transforms. As a result, convolutions
(which correspond to products in Fourier space) can be simply
expressed using shapelets. Let us consider the convolution $h(x) =
(f * g)(x)$ of 2 functions. Each function can be decomposed into
shapelet coefficients with scales $\alpha$, $\beta$ and $\gamma$,
which are then related by
\begin{equation}
\label{eq:convolution} h_{n} = \sum_{m,l=0}^{\infty} C_{nml} f_{m}
g_{l},
\end{equation}
where the convolution tensor $C_{nml}(\alpha,\beta,\gamma)$ can be
estimated analytically using a simple recursion relation
(Refregier \& Bacon 2001).

Shapelets also have simple properties under coordinate
transformations such as translation, dilation, rotation and shear.
For instance, let us consider the distortion of an image
$f({\mathbf x})$ under the action of a weak shear $\gamma_{i}$, as
occurs in weak gravitational lensing. To first order in the shear,
the distorted image can be written as
\begin{equation}
f' \simeq (1+\gamma_{i} \hat{S}_{i}) f,
\end{equation}
where $\hat{S}_{i}$ is the shear operator. It is easy to show that
this operator takes a simple form in shapelet space, namely
\begin{equation}
\label{eq:shear_op} \hat{S}_{1} = \frac{1}{2} \left(
\hat{a}_{1}^{\dagger 2} - \hat{a}_{2}^{\dagger 2} -
\hat{a}_{1}^{2} + \hat{a}_{2}^{2} \right),~~~ \hat{S}_{2} =
\hat{a}_1^{\dagger} \hat{a}_2^{\dagger} - \hat{a}_1 \hat{a}_2,
\end{equation}
where $\hat{a}^{\dagger}_{i}$ and $\hat{a}_{i}$ are the raising
and lowering operators in the QHO formalism, for each dimension
$i=1,2$. Similar operators can be constructed for the other
coordinate transformations.

\section{Measurement of Weak Lensing}
The weak distortions induced by lensing on the images of
background galaxies provide a direct measure of the distribution
of mass in the Universe. This weak lensing method is now routinely
used to study galaxy clusters, and has recently been detected in
the field (Wittman et al 2000; Bacon et al 2000; Kaiser et al
2000; Maoli et al 2001; van Waerbeke et al 2000, 2001; Rhodes et
al. 2001). Because the lensing effect is only of a few percent on
large scales, a precise method for measuring the shear is
required. The original methods of Bonnet \& Mellier (1995) and
Kaiser, Squires \& Broadhurst (KSB; 1995) are not sufficiently
accurate and stable for the upcoming weak lensing surveys (see
Bacon et al 2001, Erben et al 2001). Thus several new methods have
been proposed (Kuijken 1999; Rhodes, Refregier \& Groth 2000;
Kaiser 2000; Bernstein \& Jarvis 2001).

The remarkable properties of our basis functions make shapelets
particularly well suited for providing the basis of a new method
for measuring the shear (Refregier \& Bacon 2001). First, the
Point-Spread Function (PSF) of the instrument can be modeled by
decomposing stellar images into shapelets and by interpolating the
resulting coefficients across an image. Galaxy images can then
themselves be decomposed into shapelet coefficients, and the
analytical form of the convolution matrix (see
Eq.~[\ref{eq:convolution}]) can be used to deconvolve the
Point-Spread Function. From the properties of shapelets under
shears (eq.~[\ref{eq:shear_op}]), one can then construct a linear
estimator for the shear from the (deconvolved) coefficients of the
galaxies of the form
\begin{equation}
\label{eqn:g1g2} \widetilde{\gamma}_{i{\mathbf n}} = \frac{
f'_{\mathbf n} - \langle f_{\mathbf n} \rangle} {S_{i{\mathbf nm}}
\langle f_{\mathbf m} \rangle}
\end{equation}
where $f'_{\mathbf n}$ are the lensed coefficients, the brackets
denote an average over the unlensed galaxy population, and
${\mathbf n}$ is even (odd) for the $\gamma_{1}$($\gamma_{2}$)
component of the shear. The tensor $S_{i{\mathbf nm}}$ is the
matrix representation of the shear operators of
Equation~(\ref{eq:shear_op}). It is easy to show that these
estimators are unbiased, i.e. that $\langle
\widetilde{\gamma}_{i{\mathbf n}} \rangle = \gamma_{i}$, if the
unlensed galaxies are randomly oriented. These estimators can then
be combined to construct a minimum variance global estimator for
the shear over a sample of galaxies.

\begin{figure}[t]
\centerline{\psfig{figure=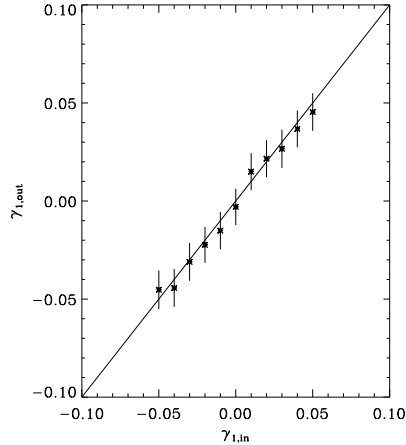,height=2.5in}} \caption{Input
shear vs recovered shear for a set of 11 simulations. Note the
linear relationship between input and recovered shear; the slope
is found to be $0.97 \pm 0.04$ for $\gamma_1$ and $1.00 \pm 0.04$
for $\gamma_2$.} \label{fig:gamma}
\end{figure}

To test the method, we used the image simulations described in
Bacon et al. (2001). These were designed to reproduce the
observational conditions of ground based telescopes such as
appropriate throughput, PSF and noise, along with the statistics
of the galaxy population observed in HST images. An artificial
shear was applied to the simulated galaxies, and the shapelet
method was used on the resulting realistic images to recover the
applied shear. The recovered shear values from the simulations are
plotted as a function of the output shear in
Figure~\ref{fig:gamma}. Clearly, the method is unbiased, and is
found to be robust with PSF shape and size. It thus provides an
improvement over the KSB method which was shown to have small but
significant instabilities and biases (Bacon et al 2001; Erben et
al. 2001).

\section{Image reconstruction with Interferometers}
Another application of shapelets is the reconstruction of images
from interferometric data, as described in detail in Chang \&
Refregier (2001). Interferometric data are collected in the $uv$
space, the Fourier-transform of the sky surface brightness. The
observed quantity is the visibility measured for each antenna pair
$(i,j)$ at time $t$ and at frequency $\nu$ and is given by
\begin{equation}
\label{eq:v_ij} V_{ij}(\nu,t) = \int d^{2}l~ \frac{A({\mathbf
l},\nu) f({\mathbf l},\nu,t)}{\sqrt{1-|l|^2}} e^{-2\pi i [ ul + vm
+w (\sqrt{1-|l|^2}-1) ]},
\end{equation}
where $f({\mathbf l},\nu,t)$ is the surface brightness of the sky
at position ${\mathbf l}=(l,m)$ with respect to the phase center,
and $A({\mathbf l},\nu)$ is the (frequency-dependent) primary
beam.

In analogy with the shapelet decomposition in real space, we wish
to decompose sources directly in the $uv$ space. However, the
discrete and finite sampling of the $uv$ plane prevents a direct
linear decomposition of the visibilities $V_{ij}$ into shapelet
coefficients. Instead, we simultaneously fit for the shapelet
coefficients of a collection of sources on the $uv$ plane, by
using a $\chi^2$ fit:
\begin{equation}
\label{eq:chi2} {\chi}^{2}=({\mathbf d} - {\mathbf M
~f})^{T}~{\mathbf C}^{-1}~({\mathbf d} - {\mathbf M~f}),
\end{equation}
where ${\mathbf d}=\{ \overline{V}_{ij} \}$ is the visibility data
vector , ${\mathbf M}=\{ \overline{V}_{ij}^{{\mathbf n}s} \}$ is
the theory matrix composed of visibilities corresponding to each
basis function, and ${\mathbf f}=\{ f_{{\mathbf n}s} \}$ is the
shapelet coefficient vector. The indices $s$ and $n$ stand for the
source number and shapelet state, respectively. The covariance
error matrix of the visibilities
\begin{equation}
{\mathbf C} = {\rm cov}[{\mathbf d},{\mathbf d}] = \left\langle
({\mathbf d} - \langle {\mathbf d} \rangle)^{T}{(\mathbf d} -
\langle {\mathbf d} \rangle) \right\rangle
\end{equation}
is provided by the interferometric hardware. The resulting
$\chi^2$-fit is linear in its parameters and can therefore be
performed by simple matrix operations.  The complex effects of
bandwidth smearing, time averaging and non-coplanarity of the
array can be fully corrected for in the process.

\begin{figure}[t]
\centerline{\psfig{figure=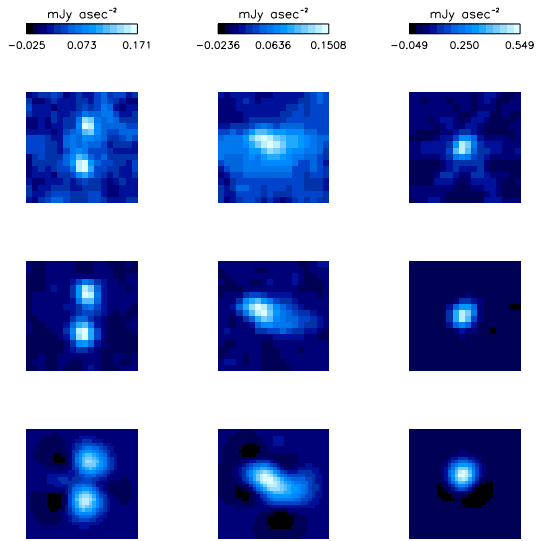,height=3.in}}
\caption{Reconstruction of three sources from one of the FIRST
grid pointings. The 'dirty' images (which are direct inverse
Fourier transformation of the data), CLEAN images, and shapelet
reconstructions are shown from top to bottom, respectively.  The
images are 32$''$ across and the resolution is about 5$''$.4
(FWHM). The dirty and CLEAN images are displayed with a 1$''$.8
pixel size, while the shapelet reconstruction images have 1$''$
pixels. For the shapelet reconstruction, we used Wiener filtering
and smoothing by a Gaussian restoring beam with a standard
deviation of 2.3$''$.} \label{fig:first}
\end{figure}

As an example we consider the observing conditions of the FIRST
radio survey (Becker et al. 1995; White et al. 1997).  Using one
of the FIRST pointings, as shown in Fig. \ref{fig:first}, we find
that our method compares well with CLEAN, the commonly used method
for interferometric imaging.  Our method has the advantage of
being linear in the shapelet parameters, of fitting all sources
simultaneously, and of providing the full covariance matrix of the
coefficients, which allows us to quantify the errors and
cross-talk in image shapes.  It is well-suited for quantitative
shape measurements which require high-precision.  In particular,
combining with the results from the previous section (Refregier \&
Bacon 2001), our results provide an accurate method for measuring
weak lensing with interferometers.

\section{Conclusions}
Shapelets provide a new method to measure the shape of galaxies.
The chosen basis functions have convenient properties under
convolutions and distortions. This makes shapelets ideally suited
for the measurement of weak lensing shear, after correcting for
the smearing effect of the point spread function. They can also be
used to reconstruct images from interferometric data. The method
is very general and is ideally suited for astrophysical
applications requiring high precision. In particular, it can be
used for image compression, modeling and deconvolution of the
point spread function, and de-projection from 2 dimensional to 3
dimensional data. Another immediate application to explore is that
of the measurement and classification galaxy morphology. Shapelets
could thus help in unifying the different observational and
theoretical studies of galaxy shapes presented in this volume.

\section*{Acknowledgments}
We thank Richard Ellis, David Helfand and Richard Massey for useful
discussions and on-going collaboration. AR was supported by a
fellowship from the EEC TMR network on gravitational lensing and by a
Wolfson College Fellowship. TC was supported by NSF grant
AST-98-0273. DJB was supported at the IfA by a PPARC postdoctoral
fellowship.

\end{document}